\documentclass[fleqn,11pt]{article}
\usepackage[cp1251]{inputenc}
\usepackage{amsmath,amssymb}
\usepackage{amsfonts,amssymb,cite}
\usepackage{graphicx}

\def\noi{\noindent}

\newcommand{\Title}[1]{\noi {{\Large\bf #1}}\\[1ex]}

\newcommand{\Author}[2]{\noi{\bf #1}\\[2ex]\noi{\normalsize\it #2}\\}

\newcommand{\Abstract}[1]{\vskip 2mm \begin{center}
        \parbox{16.4cm}{\small\noi #1} \end{center}\medskip}
\newcommand{\foom}[1]{\protect\footnotemark[#1]}

\def\nqq{\hspace*{-2em}}

\def\Jl#1#2{#1 {\bf #2},\ }

\def\ApJ#1 {\Jl{Astroph. J.}{#1}}
\def\CQG#1 {\Jl{Class. Quantum Grav.}{#1}}
\def\DAN#1 {\Jl{Dokl. AN SSSR}{#1}}
\def\GC#1 {\Jl{Grav. Cosmol.}{#1}}
\def\GRG#1 {\Jl{Gen. Rel. Grav.}{#1}}
\def\JETF#1 {\Jl{Zh. Eksp. Teor. Fiz.}{#1}}
\def\JETP#1 {\Jl{Sov. Phys. JETP}{#1}}
\def\JHEP#1 {\Jl{JHEP}{#1}}
\def\JMP#1 {\Jl{J. Math. Phys.}{#1}}
\def\NPB#1 {\Jl{Nucl. Phys. B}{#1}}
\def\NP#1 {\Jl{Nucl. Phys.}{#1}}
\def\PLA#1 {\Jl{Phys. Lett. A}{#1}}
\def\PLB#1 {\Jl{Phys. Lett. B}{#1}}
\def\PRD#1 {\Jl{Phys. Rev. D}{#1}}
\def\PRL#1 {\Jl{Phys. Rev. Lett.}{#1}}

\def\lal{&&\nqq {}}

\def\beq{\begin{equation}}
\def\eeq{\end{equation}}
\def\bear{\begin{eqnarray}}
\def\bearr{\begin{eqnarray} \lal}
\def\ear{\end{eqnarray}}
\def\earn{\nonumber \end{eqnarray}}

\newcommand{\Fig}[3]{%
\begin{center}
\parbox{8cm}{%
\refstepcounter{figure}\includegraphics[width=8cm,height=#2cm]{#1} \noindent Figure \thefigure:\quad
#3}\end{center}}
\begin{document}
\thispagestyle{empty}
\twocolumn[
%\jnumber{4}{2019}

%vspace{1cm}

\Title{Cosmological evolution of a scalar-charged degenerate cosmological plasma with Higgs scalar fields\foom 1}

\Author{Yu.G. Ignat'ev}
    {Institute of Physics, Kazan Federal University, Kremlyovskaya str., 18, Kazan, 420008, Russia}

%\Dates{June 16, 2019}{July 25, 2019}{August 1, 2019}

\Abstract
 {A mathematical model of the cosmological evolution of statistical systems of scalarly charged particles with Higgs scalar interaction is formulated and investigated. Examples are given of numerical modeling of such systems, revealing their very remarkable properties, in particular, the formation of paired bursts of cosmological acceleration.
}
\bigskip

] %%%%%%%%%%%%%%%%%%%%%%%%%%%

\section*{Intoduction}
In \cite{Ignat20}, a complete mathematical model of the cosmo\-lo\-gical evolution of the classical Higgs scalar vacuum field was formulated and studied, both by methods of qualitative analysis and numerical simulation. In such models, transitions of cosmological evolution from the stage of expansion to the stage of comp\-re\-ssion (and, conversely, for the phantom field) become possible \footnote{We do not pose in this article the task of compiling any review on a huge layer of articles devoted to cosmological models based on scalar fields. Such a brief review is contained in the cited works of the Author, in particular, \cite{YuI_Kokh19}.}. Earlier, a comprehensive study of incomplete cosmological models was carried out under the assumption that the Hubble constant is non-negative for the cases of the classical Higgs vacuum field \cite {YuI_Clas17}, the Higgs phantom field \cite{YuI_Fant17}, \cite{YuI_Agaf_Fant17} and the asymmetric scalar doublet \cite{YuI_Kokh_Russ18-19}  --  \cite{YuI_Kokh19}. If we discard a number of incorrect results of these works, which are just related to the assumption that the Hubble constant is non-negative, then one of the results of these works can be summarized as follows: at the late stages of evolution, the cosmological model always goes to inflation. The same result was confirmed by studies of the complete \cite{Ignat20} model, in which the assumption that the Hubble constant was nonnegative was removed. Thus, it can be argued that cosmological models based on vacuum scalar fields contradict the observational data on invariant cosmological acceleration in the late Universe $w <1$.

On the other hand, in a number of earlier works based on the theory of statistical systems of scalar charged particles, \cite{Ignat14_1} -- \cite{Ignat15}, in which the cosmological evolution of such systems was investigated, the possibility of four types of behavior of the corresponding cosmological models was shown, among which were models with an intermediate ultra-relativistic stage and a final non-relativistic \cite{Ignat15_2} -- \cite{Ignat17}. However, these studies were based, firstly, on an incomplete mathematical model, se\-condly, on the quadratic potential of scalar fields and, thirdly, on a scalar singlet. In this connection, the problem arises of formulating a complete mathe\-matical model of cosmological systems of scalar charged particles with Higgs scalar fields, including an asymmetric scalar doublet. Note that the\linebreak phantom scalar field due to the negativity of its kinetic energy should be considered only as part of the usual components of matter. However, as will be seen from what follows, it is the presence of a phantom field in the system that ensures the correct behavior of the cosmological model. In this article, we formulate a mathematical model of a cosmological statistical system of scalar charged particles with Higgs scalar fields, examine its basic properties and show examples of numerical modelling.

\section{Mathematical model}
\subsection{A mathematical model of a degenerate scalarly charged plasma}
The foundations of the general relativistic kinetic and statistical theory were laid in the 60s in the works of E. Tauber - J.W. Weinberg \cite{taub}, N. A. Chernikov (see, for example, \cite{chern}), A. A. Vlasov \cite{Vlasov} and others. Scalar fields in general relativistic statistics and kinetics were introduced at the beginning of the 80s in the works of the Author \cite{Ignatev1} -- \cite{Ignatev4}. Further, in \cite{Ignat14_1,Ignat14_2,Ignat15}, a mathematical model of the statistical system of scalarly charged particles was formulated, based on the microscopic description and the subsequent procedure for the transition to kinetic and hydrodynamic models. Here we will refer to the paper \ cite {Ignat15}, which contains a correct generalization of the relativistic theory both to the case of phantom scalar fields and to the sector of negative dynamic masses of scalar charged particles
\begin{equation} \label{m*}
m_*=m_a+\sum\limits_{r=1}^N  q^r_a\Phi_r,
\end{equation}
where $m_a$ is some bare particle mass of rest of the particle, which it may be zero, $\Phi_r$ is a scalar field of type $r$, $q^r$ is the scalar charge of a particle with respect to this field ($r=\overline{1,N}$).

Strict macroscopic consequences of the kinetic theory are the transport equations, including the conservation law of a certain vector current corres\-ponding to the microscopic conservation law in reactions of some fundamental charge ${\rm G}$ (if there is such a conservation law) --
\begin{equation}\label{1}
\nabla_i\sum\limits_a {\rm g}_a n^i_a=0,
\end{equation}
as well as the laws of conservation of the energy - momentum of the statistical systems:
\begin{equation}\label{2}
\nabla _{k} T_{p}^{ik} -\sum\limits_r\sigma^r\nabla ^{i} \Phi_r =0,
\mathrm{}\end{equation}
where $n^i_a$ is a numerical vector, $T^{ik}_p$ is the energy - momentum tensor (MET) of particles; $\sigma^r$ is the density of scalar charges with respect to the field $\Phi_r$ \cite{Ignat14_2}, so that
\begin{equation}\label{2a}
T^{ik}_p=\sum\limits_a T^{ik}_a; \quad \sigma^r=\sum\limits_a \sigma^r_a.
\end{equation}

Under conditions of local thermodynamic equili\-brium (LTE), the statistical system is isotropic and is described by locally equilibrium distribution functions:
\begin{equation}\label{8_0}
f^0_a={\displaystyle \frac{1}{\mathrm{e}^{(-\mu_a+(u,p))/\theta}\pm 1}},
\end{equation}
where $\mu_a$ is the chemical potential, $\theta$ is the local temperature, $u^i$ is the unit time-like vector of the dynamic velocity of the statistical system, the sign ``$+$'' corresponds to fermions, ``$-$'' -- to the bosons. Further, the kinematic momentum of the particle $p^i$ lies on the effective mass surface:
\begin{equation}\label{8_1}
(p,p)=m^2_* \Rightarrow \tilde{p}^4=\sqrt{m^2_*+\tilde{p}^2},
\end{equation}
where $\tilde{p}^{(i)}$ are the reference projections of the momentum vector, $p^2 $ is the square of the physical momentum. In this case, the macroscopic moments take the form of the corresponding moments of the ideal fluid for each of the components \cite{Ignat14_1}:
\begin{equation}\label{3}
n^i_a=n_a u^i,
\end{equation}
\begin{equation}\label{4}
T^{ik}_a=(\varepsilon_{a}+p_{a}) u^iu^k-p_{a}g^{ik},
\end{equation}
while
\begin{equation}\label{5}
(u,u)=1.
\end{equation}
The normalization relation (\ref{5}) implies the well-known identity:
\begin{equation}\label{6}
u^k_{~,i}u_k\equiv 0.
\end{equation}
Therefore, the conservation laws (\ref{2}) can be reduced to the form:
\begin{eqnarray}\label{2a}
(\varepsilon_p+p_p)u^i_{~,k}u^k=&\nonumber\\
(g^{ik}-u^iu^k)\biggl(p_{p,k}+\sum\limits_r\sigma^r\Phi_{r,k}\biggr);&\\
\label{2b}
\nabla_k[(\varepsilon_p+p_p)u^k]=
u^k\biggl(p_{p,k}+\sum\limits_r\sigma^r\Phi_{r,k}\biggr),&
\end{eqnarray}
and the law of conservation of the fundamental charge $\mathrm{G}$ \eqref{1} becomes:
\begin{equation}\label{2c}
\nabla_k n^ru^k=0,\quad n^r\equiv \sum\limits_a q^r_a n_a.
\end{equation}

Macroscopic scalars under LTE conditions have the form \cite{Ignat14_2} \footnote{To reduce the record, we omit the particle sort index in some places}:
\begin{eqnarray} \label{6a}
n_{a} =&\displaystyle \frac{2S+1}{2\pi ^{2} } m_{*}^{3} \int\limits_{0}^{\infty } \frac{{\rm sh}^{2} x{\rm ch}
xdx}{e^{-\gamma _{a} +\lambda _{*} {\rm ch} x} \pm 1} ;
\end{eqnarray}
\begin{eqnarray}
\label{6b}
\varepsilon_p =&\displaystyle \sum\limits_{a} \frac{2S+1}{2\pi ^{2} } m_{*}^{4}
\int\limits_{0}^{\infty } \frac{{\rm sh}^{2} x{\rm ch}^{2} xdx}{e^{-\gamma _{a} +
\lambda_{*} {\rm ch} x} \pm 1};
\end{eqnarray}
\begin{eqnarray}
\label{6c}
p_p =&\displaystyle \sum\limits_{a} \frac{2S+1}{6\pi ^{2} } m_{*}^{4} %
\int\limits_{0}^{\infty }\frac{{\rm sh}^{4} xdx}{e^{-\gamma _{a} +\lambda _{*} {\rm ch} x} \pm 1}; \\
\label{6d}
T_p =&\displaystyle \sum\limits_{a}\frac{2S+1}{2\pi ^{2} } m_{*}^{2} \int\limits_{0}^{\infty }
\frac{{\rm sh}^{2} xdx}{e^{-\gamma _{a} +\lambda _{*} {\rm ch} x} \pm 1};\\
\label{6e}
\sigma^r =&\displaystyle \sum\limits_{a} \frac{2S+1}{2\pi ^{2} } q^r_a m_{*}^{3}
\int\limits_{0}^{\infty } \frac{{\rm sh}^{2} xdx}{e^{-\gamma _{a} +\lambda _{*} {\rm ch} z} \pm 1} ,
\end{eqnarray}
where $T_p$ is the trace of MET particles, $\varepsilon_p=\sum\varepsilon_a$, $p_p=\sum p_a$, $\sigma^r=\sum\sigma^r_a$, $\lambda_{*}=|m _{*}|/\theta$, $\gamma_a=\mu_a/\theta$ and $S$ is the spin of particles.

Thus, under LTE conditions formally on $5 + N$ macroscopic scalar functions $\varepsilon_p, p_p, n_r $ and 3 inde\-pendent components of the velocity vector $u ^ i $ macroscopic conservation laws give $4+ N$ inde\-pendent equations (\ref{2a} ) -- (\ref {2c}) \footnote {one of the equations (\ref {2a}) is dependent on the restdue to the identity (\ref{6})}. However, not all indicated macroscopic scalars are functionally inde\-pendent, since all of them are determined by locally equi\-li\-brium distribution functions \eqref{8_0}.
At solved a series of conditions of chemical equilibrium, when only one chemical potential remains independent, the solved mass equation of the surface \footnote {See details in \cite{Ignat14_1, Ignat14_2}.}  and given scalar potentials and the scale factor  the $4 + 2r$ of macroscopic scalar $\varepsilon_p, p_p, n_r, \sigma^r$ are determined by two scalars --- some chemical potential $\mu$ and local temperature $\theta$. Thus, the system of equations (\ref{2a}) -- (\ref{2c}) turns out to be completely defined.

\subsection{Scalar fields}
In contrast to the works \cite{Ignat14_1} - \cite{Ignat15_2} in this paper we will consider Higgs scalar fields with the Lagrange function:
\begin{equation}\label{Ls}
\mathrm{L}^r_s=\frac{1}{8\pi}\biggl(\frac{\mathrm{e}_r}{2}g^{ik}\Phi_{(r),i}\Phi_{(r),k}-V_r(\Phi_r)\biggr),
\end{equation}
where the indicator $\mathrm{e}_r = +1$ for the classical scalar field and $\mathrm{e}_r=-1$ for the phantom scalar field, $V_r(\Phi_r)$ is the potential energy of the scalar field ($V=\sum_r V_r$):
\begin{equation}\label{Higgs}
V_r(\Phi_r)=-\frac{\alpha_r}{4} \left(\Phi_r^{2} -\frac{m_r^{2} }{\alpha_r} \right)^{2} ,
\end{equation}
$\alpha_r$ is the self-action constant, $m_r$ is the mass of scalar bosons, $\mathrm{L}_s=\sum_r\mathrm{L}^r_s$.

Further %
\begin{equation}\label{T_{iks}}
T^{ik}_r=\frac{1}{8\pi}\biggl(\mathrm{e}_r\Phi_r^{,i}\Phi^{,k}_r-
\frac{\mathrm{e}_r}{2}g^{ik}\Phi_{r,j}\Phi_r^{,j}+g^{ik} V_r(\Phi_r)\biggr)
\end{equation}
is the energy - momentum tensor of the $r$-th scalar field, $T^{ik}_{s} = \sum_r T^{ik}_r$. Next, we omit the constant term in the Higgs potential \eqref{Higgs}, since it leads to a simple redefinition of the cosmological constant $\lambda $.

The scalar fields $\Phi_r$ are determined by the equations for charged scalar fields with the source \cite{Ignat15} \footnote {In connection with the normalization of the Lagrange function of the scalar field, different from the normalization of the article \cite{Ignat15}, the scalar source function on the right side is multiplied by 2}:
\begin{equation}\label{Eq_Phi_r}
\mathrm{e}_r\square\Phi_r+V'_{\Phi_r}=-8\pi\sigma^r,
\end{equation}
where $\square \psi $ is the d'Alembert operator on the metric $g_{ik} $. It can be shown that, due to \eqref{2} and \eqref{Eq_Phi_r}, the conservation law for the complete MET system `` plasma + charged scalar fields '' is identical:
\begin{equation}\label{dTik=0}
\nabla_iT^{ik}=\nabla_i\bigl(T^{ik}_{p}+T^{ik}_s\bigr)\equiv0.
\end{equation}

\subsection{Unperturbed isotropic distribution}
As a background, we consider the spatially flat Friedmann metric
\begin{equation}\label{ds_0}
ds^2=dt^2-a^2(t)(dx^2+dy^2+dz^2),
\end{equation}
and as a background solution, we consider a homo\-ge\-neous isotropic distribution of matter, when all thermodynamic functions and scalar fields depend only on time. It is easy to verify that $u^i =\delta^i_4$ converts equations (\ref{2a}) into identities, and the system of equations (\ref{2b}) -- (\ref{2c}) reduces to $1 + N$ equations:
\begin{equation}\label{7a1}
\dot{\varepsilon}_p+3\frac{\dot{a}}{a}(\varepsilon_p+p_p)=\sum\limits_r\sigma^r\dot{\Phi}_r;
\end{equation}
\begin{equation}\label{7b1}
\dot{n}_r +3\frac{\dot{a}}{a}n_r=0.
\end{equation}
Thus, there remain 2 differential equations for the two thermodynamic functions $\mu$ and $\theta$. When passing to the limit $\mu\to0$ or $\theta\to0$ we get a system of two equations for one function, and the problem of the inconsistency of these equations arises,moreover, this problem does not depend on the presence of a scalar field. However, in \cite{Ignat14_2} it is shown that this problem is apparent, in fact, no contradictions in the system of equations (\ref{2a}) - (\ref{2c}) also arise in the case of a degenerate Fermi system. As it turns out, in this case, the laws of conservation of charge \eqref{7b1} are a direct consequence of the law of conservation of energy \eqref{7a1}.

We first consider a one-component statistical system of scalarly charged fermions under conditions of complete degeneracy:
\begin{equation}\label{2_1}
\theta\to 0.
\end{equation}
when the locally equilibrium distribution function of fermions \eqref{8_0} takes the form of a step function \cite{Ignat14_2}:
\begin{equation}\label{2_2}
f^0(x,P)=\chi_+(\mu-\sqrt{m_*^2+p^2}),
\end{equation}
where $\chi_+ (z)$ is the Heaviside step function. In this case, however, we will admit the presence of several scalar fields with respect to which the same particle can have different scalar charges $q^r$.

The result of integrating macroscopic densities \eqref{6a} -- \eqref{6e} with respect to the distribution \eqref{2_2} expressed in elementary functions \cite{Ignat14_2}:
\begin{equation}\label{2_3}
n=\frac{1}{\pi^2}p_F^3;
\end{equation}
\begin{equation}\label{2_3a}
{\displaystyle
\begin{array}{l}
\varepsilon_p = {\displaystyle\frac{m_*^4}{8\pi^2}}F_2(\psi);
\end{array}}
\end{equation}
\begin{equation}\label{2_3b}{\displaystyle
\begin{array}{l}
p_p ={\displaystyle\frac{m_*^4}{24\pi^2}}(F_2(\psi)-4F_1(\psi))
\end{array}}
\end{equation}
\begin{equation}\label{2_3c}{\displaystyle
\begin{array}{l}
\sigma^r={\displaystyle\frac{q^r\cdot m_*^3}{2\pi^2}}F_1(\psi),
\end{array}}
\end{equation}
where the dimensionless function is introduced $\psi$
\begin{equation}\label{psi}\psi=p_F/m_*,
\end{equation}
equal to the ratio of the Fermi momentum $p_F$ to the effective mass of the fermion, and to reduce writing, the functions
$F_1(\ psi)$ and  $F_2(\ psi)$ were introduced:
\begin{equation}\label{F_1}
F_1(\psi)=\psi\sqrt{1+\psi^2}-\ln(\psi+\sqrt{1+\psi^2});
\end{equation}
\begin{equation}
\label{F_2}
F_2(\psi)=\psi\sqrt{1+\psi^2}(1+2\psi^2)-\ln(\psi+\sqrt{1+\psi^2}).
\end{equation}
The functions $ F_1 (x) $ and $ F_2 (x) $, firstly, are odd:
\begin{equation}\label{F(-x)}
F_1(-x)=-F_1(x);\quad F_2(-x)=-F_2(x),
\end{equation}
and, secondly, they have the following asymptotics:
\begin{eqnarray}\label{F,x->0}
\left.F_1(x)\right|_{x\to 0}\simeq \frac{2}{3}x^3; \left.F_2(x)\right|_{x\to 0}\simeq \frac{8}{3}x^3;\nonumber\\
\left.(F_2(x)-4F_1(x))\right|_{x\to 0}\simeq \frac{8}{5}x^5;
\end{eqnarray}
\begin{eqnarray}\label{F,x->8}
\left.F_1(x)\right|_{x\to\infty}\simeq x|x|;\quad \left.F_2(x)\right|_{x\to\infty}\simeq 2x^3|x|.
\end{eqnarray}
It is easy to verify the validity of the identity:
\begin{equation}\label{E_P_f}
\varepsilon_p+p_p\equiv \frac{m^4_*}{3\pi^2}\psi^3\sqrt{1+\psi^2}.
\end{equation}
Note that in the extended theory \cite{Ignat15}, the effective mass \eqref{m *} can also be a negative quantity. The requirement of symmetry between particles and antiparticles ($ q \to -q $) leads to the condition that the seed mass is equal to zero in the formula \eqref {m *}.
Therefore, for the effective particle mass we have:
\begin{equation}\label{m_*(pm)}
m_*=q\Phi.
\end{equation}
Moreover, we do not exclude the possibility of a negative effective mass of fermions, in particular, the effective masses of particles and antiparticles in this case will differ in sign, since $\overline {q} = - q$. Note that the effective mass of particles is not a heavy mass, which, in contrast to the effective mass, is determined by the total energy, i.e., $m=\sqrt{m_*^2+p^2}$,  so $m(p = 0)=|m_*|$ (see \cite{Ignat15}). So, due to $\psi(-\Phi)= -\psi(\Phi)$ and the oddness properties \eqref{F(-x)} of the formula \eqref{2_3a} -- \eqref{2_3c}, the following trans\-formation laws
\begin{eqnarray}\label{trans_eps}
\varepsilon_p(-\Phi)=-\varepsilon_p(-\Phi);\nonumber\\
p_p(-\Phi)=-p_p(\Phi);\, \sigma(-|\Phi)=\sigma(\Phi).\
\end{eqnarray}
We also note that in the case of a seemingly more standard version of $m_* =|q\Phi|\geqslant 0$, compatibility problems for the basic equations arise.

Further, the MET of the scalar field in the unperturbed state also takes the form MET of an ideal isotropic fluid:
\begin{equation} \label{MET_s}
T_{s}^{ik} =(\varepsilon_s +p_{s} )u^{i} u^{k} -p_s g^{ik} ,
\end{equation}
moreover:
\begin{eqnarray}\label{Es}
\varepsilon_s=\frac{1}{8\pi}\sum\limits_r\biggl(\frac{\mathrm{e}_r}{2}\dot\Phi_r^2+V_r(\Phi_r)\biggr);\\
\label{Ps}
p_{s}=\frac{1}{8\pi}\sum\limits_r\biggl(\frac{\mathrm{e}_r}{2}\dot\Phi_r^2-V_r(\Phi_r)\biggr),
\end{eqnarray}
so that:
\begin{equation}\label{e+p}
\varepsilon_s+p_{s}=\frac{1}{8\pi}\sum\limits_r\mathrm{e}_r\dot{\Phi}_r^2.
\end{equation}
The equation of scalar fields in the Friedmann metric takes the form:
\begin{equation}\label{Eq_S_t}
\mathrm{e}_r\biggl(\ddot{\Phi}_{r}+3\frac{\dot{a}}{a}\dot{\Phi}_{r}\biggr)+m^2_r\Phi_r-\alpha_r\Phi^3_r= -8\pi\sigma^r(t),
\end{equation}
where $(r=\overline{1,}N)$.

\subsection{Complete system of background\\ equations}
We consider the standard Einstein equations with the $\Lambda$ - term:
\begin{equation}\label{EqEinst}
G^i_k\equiv R^i_k-\frac{1}{2}R\delta^i_k=8\pi T^i_k+\Lambda\delta^i_k.
\end{equation}
We write Einstein's independent background equations for the Friedmann metric \eqref{ds_0}:
\begin{equation}\label{Einstein^1_1}
2\frac{\ddot{a}}{a}+\frac{\dot{a}^2}{a^2}+\sum\limits_r\biggl(\frac{e_r\dot{\Phi_r^2}}{2}
-\frac{m_r^2\Phi_r^2}{2}+\frac{\alpha_r\Phi_r^4}{4}\biggr)+8\pi p_p=\Lambda;
\end{equation}
\begin{equation}\label{Einstein^4_4}
3\frac{\dot{a}^2}{a^2}-\sum\limits_r\biggl(\frac{e_r\dot{\Phi_r^2}}{2}
+\frac{m_r^2\Phi_r^2}{2}-\frac{\alpha_r\Phi_r^4}{4}\biggr)-8\pi\varepsilon_p=\Lambda.
\end{equation}
Due to the energy - momentum conservation law \eqref{dTik=0}, \eqref{7a1} of the field equations \eqref{Eq_S_t} one of the Einstein equations \eqref{Einstein^1_1} - \eqref{Einstein^4_4} is differentially algeb\-raic consequence of the remaining equations. In \cite{Ignat20} shows that to study a dynamical system it is more con\-venient to consider their difference instead of these Einstein equations, taking into account the identity for the Hubble constant $H =\dot{a}/a$, --
\[ \dot{H}\equiv \frac{\ddot{a}}{a}-\frac{\dot{a}^2}{a^2}.\]
Thus, we obtain the necessary equation:
\begin{equation}\label{(11-44)-0}
\dot{H}+4\pi(\varepsilon+p)=0,
\end{equation}
where $\varepsilon=\varepsilon_p+\varepsilon_s$ and $p=p_p+p_s$.

Further, according to \cite{Ignat20}, we introduce the total energy, $\mathcal {E}$, of cosmological matter:
\begin{equation}\label{E}
\mathcal{E}=\frac{1}{8\pi}(3H^2-\Lambda)-\varepsilon,
\end{equation}
with the help of which the Einstein equation \eqref{Einstein^4_4} can be given a simple form:
\begin{equation}\label{E=0}
\mathcal{E}=0,
\end{equation}
reflecting the fact that the total energy of the spatially flat Friedman universe is zero.

Differentiating in time the total energy \eqref{E} taking into account the field equations \eqref{Eq_S_t} and the relations \eqref{7a1}, \eqref{Es}, \eqref{e+p} and \eqref{(11-44)-0}, we obtain the energy conservation law
\begin{equation}\label{dE/dt=0}
\frac{d}{dt}\mathcal{E}=0\Rightarrow \mathcal{E}=\mathcal{E}_0.
\end{equation}
Thus, the consequence of the considered system of dynamic equations \eqref{7a1}, \eqref {Eq_S_t} and \eqref{(11-44)-0} is the law of conservation of the total energy of the cosmological system \eqref{dE/dt=0} $\mathcal{E}=\mathcal{E}_0$. the Einstein equation \eqref{Einstein^4_4} is a particular integral of this system $\mathcal{E}_0=0$. A similar situation arises for the vacuum scalar fields \cite{Ignat20}. This means that the first integral \eqref{E=0} can be considered as the initial condition in the Cauchy problem for the cosmological model.

In particular, for a degenerate Fermi system, taking into account \eqref{E_P_f} we obtain from \eqref{(11-44)-0} the equation:
\begin{equation}\label{11-44}
\dot{H}+\sum\limits_r \frac{e_r\dot{\Phi}^2_r}{2}+\frac{4}{3\pi}m^4_*\psi^3\sqrt{1+\psi^2}=0.
\end{equation}

The system of equations \eqref{7a1}, \eqref{Eq_S_t} and \eqref{11-44} together with the definitions \eqref{2_3a} -- \eqref{2_3c} describes a closed mathematical model of the cosmological evolution of a completely degenerate Fermi system with scalar interaction .

Differentiating the energy density of the Fermi system (\ref{2_3a}) taking into account the identity \eqref{E_P_f} we bring the energy conservation law for the Fermi system (\ref{7a1}) to the form of the equation:
\begin{equation}\label{Eq_Pl}
\frac{d}{dt}\ln m_*\psi a=0.
\end{equation}
From this, taking into account the definition of the function $\psi$ (\ref{psi}) we get:
\begin{equation}\label{ap}
ap_F={\rm Const}.
\end{equation}
From here, taking into account (\ref{3}) we obtain the law of conservation of the number of fermions (see \cite{Ignat14_2}):
\begin{equation}\label{na3}
a^3n={\rm Const}.
\end{equation}
Thus, despite the apparent complexity of the equation (\ref{7a1}),  its solution is easy to find - from the law of conservation of energy of the Fermi system, the law of conservation of the number of particles is obtained. We can say that the law of conservation of the scalar charge in the form \eqref{2c} is, at least in our case, redundant. note that, unlike the law of conservation of electric charge, this law does not follow from anywhere, but, nevertheless, is fulfilled.

Following \cite{Ignat20} we also introduce a nonnegative effective energy of the cosmological system $\mathcal{E}_{eff}$ according to \eqref{E} and \eqref{E=0}
\begin{eqnarray}\label{E_eff}
\mathcal{E}_{eff}=\varepsilon+\frac{\Lambda}{8\pi}\geqslant 0\Rightarrow \nonumber\\
\frac{\mathrm{e}}{2}\dot\Phi^2+\frac{m^2\Phi^2}{2}-\frac{\alpha\Phi^4}{4}+\frac{m_*^4}{\pi}F_2(\psi)+\Lambda\geqslant 0.
\end{eqnarray}

\section{Dynamic system analysis}
\subsection{A dynamical system for a one -\\ component degenerate cosmological plasma}
In this article, we will consider a cosmological model based on a one-component degenerate Fermi system and a single scalar field $\Phi$. An important circumstance is that the energy conservation law of the statistical system \eqref{7a1} for a degenerate Fermi system is completely equivalent to the equation \eqref{Eq_Pl}. Instead of choosing the non-negative dynamic variable $a(t)\geqslant0$, choosing $\xi(t)$
\begin{eqnarray}\label{dxi/dt}
\xi=\ln a,\quad \xi\in(-\infty,+\infty);&\\
\label{dxi/dt}
\dot{\xi}=H,&
\end{eqnarray}
we use the integral \eqref{ap} of the energy conservation law \eqref{Eq_Pl} to find the function $\psi(t)$ \eqref{psi}
\begin{equation}\label{psi0}
\psi=\frac{p^0_f \mathrm{e}^{-\xi}}{m_*}.
\end{equation}
Thus, taking into account \eqref{m_*(pm)} we obtain the expression for the function $\psi(t)$:
\begin{equation}\label{psi1}
\psi=\frac{p^0_f \mathrm{e}^{-\xi}}{q\Phi}\equiv \frac{\beta}{\Phi}\mathrm{e}^{-\xi},
\end{equation}
where
\begin{equation}\label{beta}
\beta=\frac{p^0_f}{q};\quad p^0_f=p_f(\xi=0).
\end{equation}
Further, for the scalar charge density \eqref{2_3c} we obtain the expression
\begin{equation}\label{sigma}
\sigma={\displaystyle\frac{q^4\Phi^3}{2\pi^2}}F_1(\psi).
\end{equation}
Assuming further
\begin{equation}\label{dPhi/dt}
\dot{\Phi}=Z,
\end{equation}
we write the field equation in these notations \eqref{Eq_S_t}
\begin{eqnarray}\label{dZ/dt}
\dot{Z}=-3HZ-e m^2\Phi +e\Phi^3\biggl(\alpha-\frac{4q^4}{\pi}F_1(\psi)\biggr).
\end{eqnarray}
Further, the equation \eqref{11-44} takes the form:
\begin{equation}\label{dH/dt_0}
\dot{H}=- \frac{eZ^2}{2}+\frac{4}{3\pi}q^4\Phi^4\psi^3\sqrt{1+\psi^2}.
\end{equation}
Moreover, the first integral of the system of equations \eqref{E=0} takes the form:
\begin{eqnarray}\label{Surf_Einst}
\Sigma_E:\;3H^2-\Lambda-\frac{q^4\Phi^4}{\pi}F_2(\psi)-\nonumber\\
\frac{eZ^2}{2}-\frac{m^2\Phi^2}{2}+\frac{\alpha\Phi^4}{4}=0.
\end{eqnarray}
The equation \eqref{Surf_Einst} is an algebraic equation for the dynamic variables $\Phi, \xi, Z, H $ and describes somea hypersurface in the arithmetic space $\mathbb{R}_4=\{\Phi,\xi,$ $Z,H\}$, which we will call \cite{Ignat20} \emph{Einstein hypersurface}. All phase trajectories of the dynamical system \eqref {dPhi/dt}, \eqref{dxi/dt}, \eqref{dZ/dt} and \eqref{dH/dt}, as well as the starting points, must lie on Einstein hypersurface. Since \eqref{Surf_Einst} is the first integral of a dynamical system, for to solve the Cauchy problem, it suffices to require that the initial point of the dynamic trajectory of the Einstein hypersurface belong to.

Further, the points of the phase space $\mathbb {R}_4$, at which the effective energy \eqref{E_eff} is negative, are not available for the dynamical system.These points lie on the hypersurface of the phase space $S_{E} \subset \mathbb {R}_4 $, which is a cylinder with the axis $OH $:
\begin{equation}\label{S_E}
S_E:\; \Lambda+\frac{q^4\Phi^4}{\pi}F_2(\psi)+
\frac{eZ^2}{2}+\frac{m^2\Phi^2}{2}-\frac{\alpha\Phi^4}{4}=0,
\end{equation}
moreover, the hypersurface of zero effective energy \eqref{S_E} touches the Einstein hypersurface \eqref{Surf_Einst} in the hyperplane
$H=0$:
\begin{equation}\label{H=0}
\Sigma_E \cap S_E =H=0.
\end{equation}

Further, as can be seen from the equation \eqref{dH/dt_0} in the case of a scalar neutral statistical system ($q\equiv0$), the sign of the derivative of the Hubble constant is completely determined by the indicator $e$: for a classical scalar field ($e=+1 $) $\dot{H}<0$, and for a phantom scalar field ($e=-1 $) always $\dot{H}> 0 $. The play of these factors during cosmological evolution can also fine-tune the model parameters to ensure the desired behavior. In the presence of charged matter, its contribution to this game, as can be seen from \eqref{dH/dt_0}, is determined by the sign of the scalar potential: for $\Phi>0$ it contributes to an increase in the Hubble constant, for $\Phi<0$ it decreases. It should be noted that with a suitable Einstein hypersurface topology, cosmological models based on single vacuum scalar fields at the final stage of evolution will go either to the inflation compression mode (classical field) or to the inflation expansion mode (phantom field).

Note that instead of the equation \eqref{dH/dt_0} we can consider the equivalent equation (see \cite{Ignat20}), subs\-ti\-tuting the expression for $Z^2$ from \eqref{Surf_Einst} in \eqref{dH/dt_0}:
\begin{eqnarray}\label{dH/dt}
\dot{H}=-3H^2+\Lambda+\frac{eq^4\Phi^4}{\pi}F_2(\psi)+\nonumber\\
-\frac{m^2\Phi^2}{2}+\frac{\alpha\Phi^4}{4}+\frac{4}{3\pi}q^4\Phi^4\psi^3\sqrt{1+\psi^2}.
\end{eqnarray}
\subsection{Singular points of a dynamical system}
The singular points of a dynamical system repre\-sented by a normal autonomous system of differential equations are determined by algebraic equations obtained by equating to zero the derivatives of all dynamic variables. Thus, from \eqref{dxi/dt}, \eqref{dPhi/dt}, \eqref{dZ/dt} and \eqref{dH/dt} we obtain the system of algebraic equations for finding the coordinates of the singular points:
\begin{eqnarray}
\label{Z=0}
Z=0;\\
\label{H}
H=0;\\
\label{-3HZ}
\Phi^3\biggl(\alpha-\frac{4q^4}{\pi}F_1(\psi)\biggr)-m^2\Phi=0;\\
\label{dotH=0}
\frac{4}{3\pi}q^4\Phi^4\psi^3\sqrt{1+\psi^2}=0.
\end{eqnarray}
In addition, we must take into account the integral of the total energy \eqref{Surf_Einst}, -- the coordinates of the singular point must satisfy this equation, which given \eqref{Z=0}, takes the form:
\begin{equation}\label{Surf_0}
\Lambda+\frac{eq^4\Phi^4}{\pi}F_2(\psi)-\frac{m^2\Phi^2}{2}+\frac{\alpha\Phi^4}{4}=0.
\end{equation}

From \eqref{dotH=0} it follows 1. $\psi=0$ or 2. $\Phi=0$. We first investigate the first possibility $\psi =0$. Since $F_1(0) 0$, according to \eqref{-3HZ} we get the equation on $\Phi $
\begin{equation}\label{alpha=q^2}
\alpha\Phi^3-m^2\Phi=0,
\end{equation}
where do we get the roots from:
\begin{equation}\label{Phi_0}
\Phi_0=0;\quad \Phi_\pm=\pm \sqrt{\frac{m^2}{\alpha}}.
\end{equation}
In this case, the remaining equation \eqref{Surf_0} gives the relationship between the fundamental constants
\begin{equation} \label{Lambda0}
\Lambda=\Lambda_0\equiv\frac{m^4}{4\alpha},
\end{equation}
at which there exist a singular point $M_\Phi$
\begin{equation}\label{M_Phi}
\!\!\! M_\Phi^\pm: \; \biggl(\pm\sqrt{\frac{m^2}{\alpha}},+\infty,0,0\biggr), \; (\Lambda=\frac{m^4}{4\alpha}>0).
\end{equation}
This case corresponds to the singular points of the vacuum scalar field without charged fermions \cite{Ignat20}, and the cosmological constant $\Lambda_0$ is fully generated by the Higgs field.

We are now investigating the second possibility $\Phi=0$. In this case, the equation \eqref{-3HZ} becomes an identity, and the equation  \eqref{Surf_0} gives $\Lambda=0$. What about $\psi\to\pm \infty$ and the dynamic variable $\xi_0$ can take any values:
\begin{equation}\label{M_xi}
M_\xi^\pm: \quad \biggl(\xi_0,0,0,0\biggr), \quad (\forall\xi_0,\; \Lambda=0).
\end{equation}
Calculating the basic matrix of the dynamical system \eqref{dPhi/dt}, \eqref{dxi/dt}, \eqref{dZ/dt} and \eqref{dH/dt}
\[A=\left|\left|\frac{\partial X_i}{\partial x_k}\right|\right|,\]
it is easy to show that in both cases - \eqref{M_Phi} and \eqref{M_xi} this matrix is degenerate, therefore a qualitative theory differential equations for our dynamic system, in contrast to a dynamic system with vacuum scalar fields, does not give anything. Let us therefore proceed to numerical modeling.

\section{Numerical modeling and\\ discussion of results}
First, we note that the topology of the Einstein hypersurface \eqref{Surf_Einst} can be quite complex, and at the same time, the phase trajectories of the system can be complex, since they lie on this hypersurface. Bearing in mind a wide variety of dynamic system behavior models depending on the fundamental parameters of the model $\mathbf{P}=[e,\alpha,\beta,m,q,\Lambda], $ in this article we restrict ourselves to the study of some special cases, relating full study and general conclusions to a more detailed article, which we hope to present in the near future. In the future, for brevity, we will describe the initial conditions for our model with the ordered set $\mathbf{I}=[\Phi_0,\xi_0,Z_0,\epsilon],$ where the indicator $\epsilon=\pm1$ takes the value $+1$, if the initial state of the dynamic system corresponds to the expansion phase $H_0>0 $, and the value $ -1 $, if the initial state of the dynamic system corresponds to the compression phase $H_0<0$. Recall that the initial value of the Hubble constant is determined from the equation \eqref{Surf_Einst}.

\subsection{An example of a model with a classic scalar field}
 In Fig. \ref{Fig1} -- \ref {Fig2} shows an example of a phase trajectory of a cosmological system based on a classical scalar field, with parameters $\mathbf {P}=[1,0.1,0.1,1,0.001,0.01]$ and initial conditions $\mathbf{I}=[0.1,1,0.1,\pm1]$.
\Fig{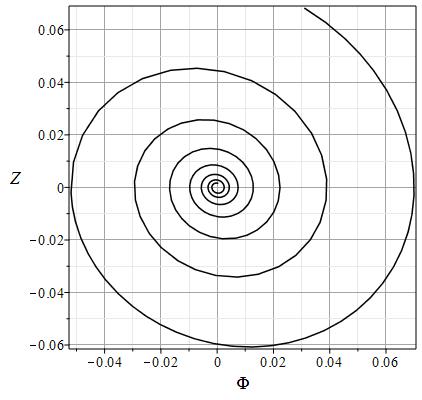}{7}{\label{Fig1}Phase trajectory of a model with a classical scalar field in the plane $\{\Phi, Z\}$; $H_0>0$.}
Note that the topology of the Einstein surface in the case under consideration makes it impossible for the phase trajectory to transition from the region $H>0$ to the region $H<0$. Therefore, the phase trajectories in the upper half-plane $H>0$ are collected at the minimum point $H_{min}>0$, and the trajectories in the lower half-plane, on the contrary, leave the maximum point $H_{max}<0$ and go ``down'' unlimitedly.
\Fig{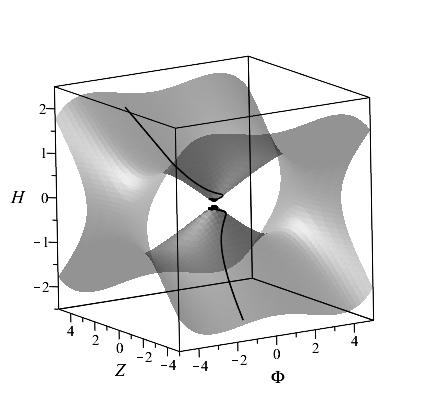}{7}{\label{Fig2}Phase trajectories of a model with a classical scalar field on an Einstein surface in a three-dimensional section $\{\Phi,Z,H\}$. In the upper part of Einstein's surface, the trajectories roll down to the lower point, and in the lower part roll down from the top.}
\Fig{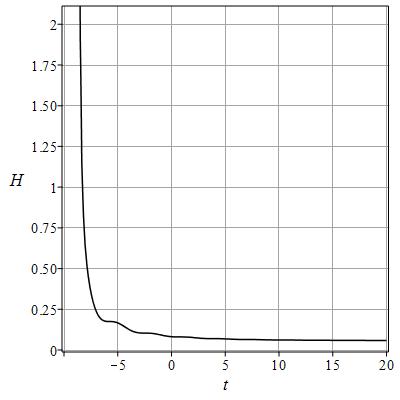}{7}{\label{Fig3}The evolution of the Hubble constant $H(t)$ in the system under study.}

In Fig. \ref{Fig3} - \ref{Fig6} shows the behavior of the basic physical parameters of the model under study with the classical Higgs scalar field: Hubble constant $H$, effective energy $\mathcal{E}_{eff}$ \eqref{E_eff}, invariant cosmological acceleration $w$
\begin{equation}\label{w}
w=1+\frac{\dot{H}}{H^2}
\end{equation}
and invariant curvature
\[\sigma=\sqrt{R_{ijkl}R^{ijkl}}=H^2\sqrt{6(1+w^2)}.\]
\Fig{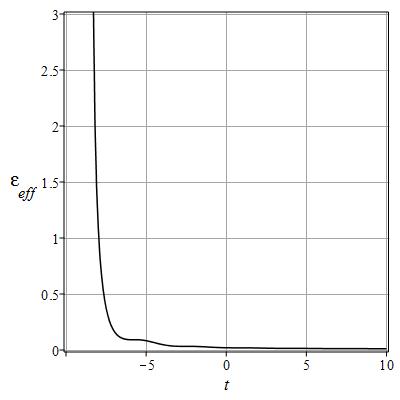}{7}{\label{Fig4}Evolution of the effective energy $\mathcal {E}_{eff}(t)$ in the system under study.}
\Fig{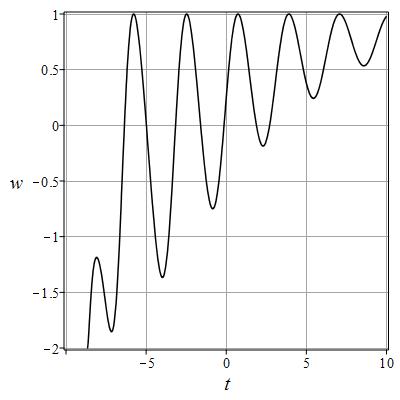}{7}{\label{Fig5}The evolution of invariant cosmological acceleration $w$ in the system under study.}
\subsection{Phantom scalar field model example}
%
%[-1,0.1,0.1,1,0.001,0.01],[0.1,1,0.1,1]
In Fig. \ref{Fig7} -- \ref {Fig12} shows the results of numerical simulations for a cosmological statistical system with phantom interaction with parameters $\mathbf{P} = [- 1,0.1,0.1,1,$  $0.001,0.01]$ and initial conditions $\mathbf{I}=[0.1,1,0.1, \pm1]$.
\Fig{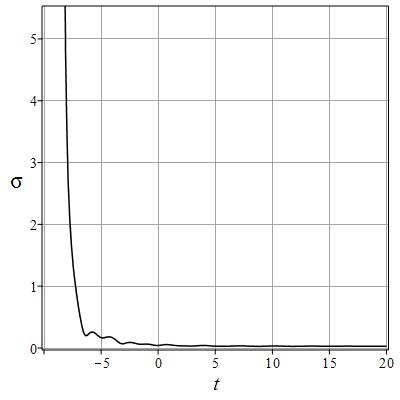}{7}{\label{Fig6}Evolution of the invariant curvature of $\sigma(t)$ in the system under study.}
%}
%

In Fig. \ref{Fig8} --  \ref{Fig9} shows the phase trajectories of the model with a phantom scalar field on the Einstein surface in a three - dimensional section $\{\Phi,Z,H\}$, and the first case corresponds to $\epsilon=+1$ and the second - $\epsilon=-1$. We see how in the first case the trajectory rises from the lower part of the left cavity of the Einstein surface to its upper point, starting from the points of the neck, and in the second case, the trajectory rises from the lower part of the right cavity of the Einstein surface to the upper point of the left plane, slipping through the neck.

\Fig{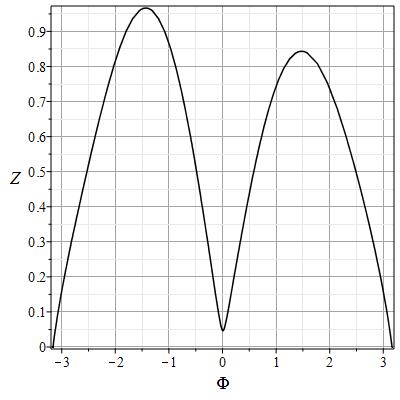}{7}{\label{Fig7}Phase trajectory of a model with a phantom scalar field in the plane $\{\Phi,Z\}$; $H_0>0$.}
Despite the seeming exoticism, the overgiant acceleration bursts are not dangerous for the cosmological model, since they pass under conditions of zero velocity of cosmological expansion $H\to0$ and instantly flat space - time $\sigma\to 0$. Nevertheless, these possible bursts are of certain interest for observational cosmology, so they can give evidence of a change in the compression mode to the expansion mode in the early Universe.
\Fig{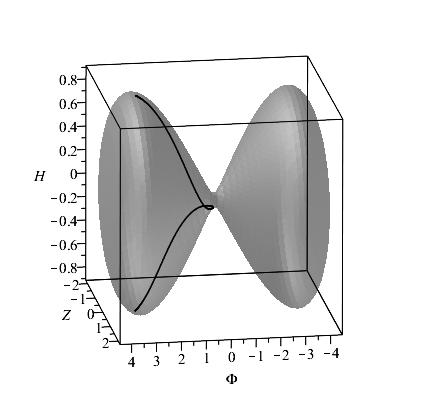}{7}{\label{Fig8}Phase trajectory of the phantom field, $\epsilon=+1$.}
We will especially dwell on the behavior of the invariant cosmological acceleration $w$ for a statistical system with a phantom field (see Fig. \ref{Fig11} and Fig. \ref{Fig12}). Here we observe two giant bursts of acceleration for a system that started from the expansion phase (Fig. \ref{Fig11}) and the compression phase (Fig. \ref {Fig12}), and the graphs in these figures are, in fact, a mirror image of each other.

In Fig. \ref{Fig11} the giant surge of acceleration $w\sim 10^2$ precedes the supergiant $ w\to+\infty$, and in Fig. \ref{Fig12} the burst sequence changes. First, we note that overgig acceleration bursts $w\to+\infty$ are associated with the passage of the point $H=0$ (see the formula \eqref{w}) and are not observed in models with a quadratic interaction potential. Secondly, the giant bursts of $w\sim10^2$ are characteristic of a theory with the quadratic potential \cite{Ignat15_2} -- \cite{Ignat17}.
\Fig{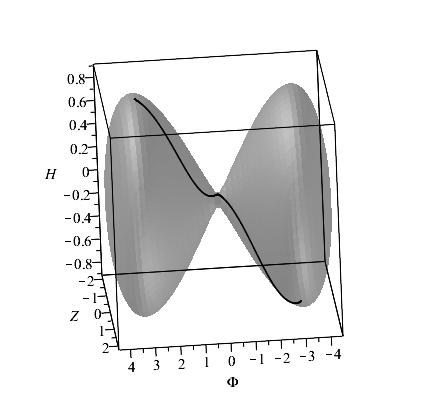}{7}{\label{Fig9}Phase trajectory of the phantom field, $\epsilon=-1$.}
\Fig{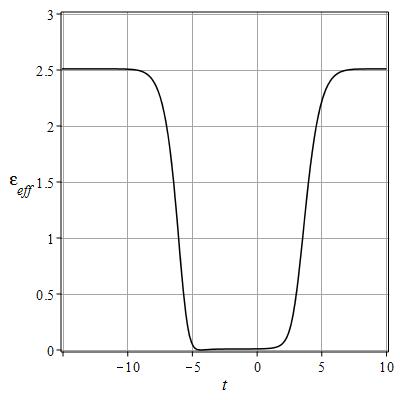}{7}{\label{Fig10}Evolution of the effective energy $\mathcal {E}_{eff}(t)$ in the system under study.}

In the near future, we intend to publish more detailed studies of the models of cosmological evolution of scalarly charged statistical systems with Higgs scalar fields.

\subsection*{Funding}

 This work was funded by the subsidy allocated to Kazan Federal University for the
 state assignment in the sphere of scientific activities.

\Fig{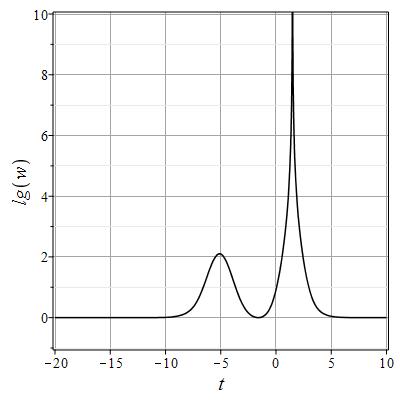}{7}{\label{Fig11}The evolution of invariant cosmological acceleration $w$ in the system under study $\epsilon=+1$.}
\Fig{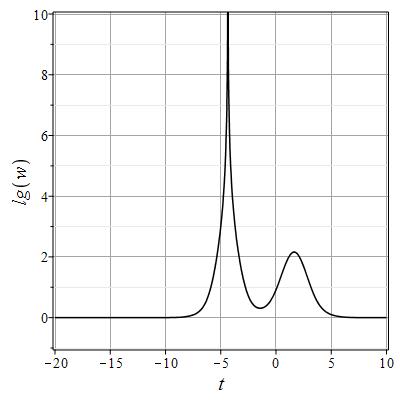}{7}{\label{Fig12}The evolution of invariant cosmological acceleration $w$ in the system under study $\epsilon=-1$.}

\end{document}